\documentclass[a4paper,12pt]{article}
\usepackage{graphicx}
\usepackage{cite}
\usepackage{amssymb,amsmath,latexsym,enumerate}
\usepackage{subfigure,epsfig}
\usepackage{array,longtable,lscape,bigints}
\usepackage{color}
\usepackage{enumerate}
\usepackage{authblk}

\begin{document}

\title{Emergence of collective self-oscillations in  minimal lattice models with feedback}

\author[1,2]{Dmitry Sinelschikov}
\author[3,4]{Anna Poggialini}
\author[5,6]{Maria Francesca Abbate}
\author[1,7]{Daniele De Martino}

\affil[1]{\footnotesize{Biofisika Institutua (UPV/EHU, CSIC) and Fundaci\'on Biof\'isica Bizkaia, Leioa E-48940, Spain.}}
\affil[2]{HSE University, 34 Tallinskaya Street, 123458, Moscow, Russian Federation}
\affil[3]{ Dipartimento di Fisica, Sapienza Universit\`a di Roma, P.le A. Moro, 2, I-00185 Rome, Italy.}
\affil[4]{ `Enrico Fermi' Research Center (CREF), Via Panisperna 89A, 00184 - Rome, Italy}
\affil[5]{ Digital Biologics Platform (DBxP) Site Lead France, Large Mol. Res. Platform at Sanofi}
\affil[6]{Laboratoire de physique de l’École normale supérieure, CNRS, PSL University, Sorbonne Université, and Université de Paris, 24 rue Lhomond, 75005 Paris, France}
\affil[7]{ Ikerbasque Foundation, Bilbao 48013, Spain.}

\date{}

\maketitle

\begin{abstract}
The emergence of collective oscillations and synchronization is a widespread phenomenon in complex systems. While widely studied in the setting of dynamical systems, this phenomenon is not well understood in the context of out-of-equilibrium phase transitions in many body systems. Here we consider three classical lattice models, namely the Ising, the Blume--Capel and the Potts models, provided with a feedback among the order and control parameters. With the help of the linear response theory we derive low-dimensional nonlinear dynamical systems for mean field cases. These dynamical systems  quantitatively reproduce many-body stochastic simulations.
In general, we find that the usual equilibrium phase transitions are taken over by more complex bifurcations where nonlinear collective self-oscillations emerge, a behavior that we illustrate by the feedback Landau theory. For the case of the Ising model, we obtain that the bifurcation that takes over the critical point is non-trivial in finite dimensions. Namely, we provide numerical evidence that in two dimensions the most probable value of a cycle's amplitude follows the Onsager law for slow feedback. We illustrate multi-stability for the case of discontinuously emerging oscillations in the Blume--Capel model, whose tricritical point is substituted by the Bautin bifurcation. For the Potts model with $q=3$ colors we highlight the appearance of two mirror stable limit cycles at a bifurcation line and characterize the onset of chaotic oscillations that emerge at low temperature through either the Feigenbaum cascade of period doubling or the Aifraimovich--Shilnikov scenario of a torus destruction. We also show that  entropy production singularities as a function of the temperature correspond to qualitative change in the spectrum of Lyapunov exponents. Our results show that mean-field collective behaviour can be described by the bifurcation theory of low-dimensional dynamical systems, which paves the way for the definition of universality classes of collective oscillations.
\end{abstract}

\section{Introduction}
The emergence of oscillations in complex systems is an widespread phenomenon that appear across various fields of science. For example, oscillatory behavior can arise in biological systems, chemical reactions and mechanical systems (see, e.g. \cite{andronov1966theory,strogatz2018nonlinear,Murray2002}). These oscillations often manifest themselves as collective behavior, where individual components interact and synchronize their activities over time, and studying underlying mechanisms is an important problem for understanding of complex systems.

One theoretical framework that could potentially shed light on the emergence of oscillations is the one of non-equilibrium phase transitions in statistical physics. While a deep and powerful physical theory underlies the classification of equilibrium critical points, this same task has not been yet achieved for out-of-equilibrium phase transitions, in particular for the emergence of collective oscillations and synchronization phenomena.
After the work of Landau and co-workers \cite{landau1969statistical} it was realized that the plethora of experimental data on continuous phase transitions could be unified on the basis of the symmetries of the underlying interacting degrees of freedom. One of  paradigmatic examples is the equivalence of critical exponents for liquid-gas and paramagnetic-ferromagnetic phase transitions \cite{landau1969statistical}. This led to the concept of universality classes \cite{pokrovskii1979fluctuation}, that is connected to field theories and the renormalization group \cite{cardy1996scaling}.

Recent works try to establish general and universal concepts in out-of-equilibrium phenomena: example ranges from the  proposal of the directed percolation universality class, including reaction-diffusion and epidemic spreading \cite{hinrichsen2000non} to the modeling of non-reciprocal phase transitions \cite{fruchart2021non}. Moreover, studies of synchronization phenomena in many body systems stands out as a very active area of research taking into account its importance for modeling complex systems \cite{gambuzza2021stability, spelat2022dual, millan2019synchronization}. One of the difficulties related to out-of-equilibrium system is a lack of general variational principles and reference free energy landscapes.
Consequently, a great majority of the studied models are variations of the Kuramoto model \cite{kuramoto1975self, sone2022topological,carletti2023global,millan2020explosive, o2016dynamics}, where the interacting units are postulated as oscillators at the outset and their phase coherence is analyzed. Dynamical phase transitions in classical lattice models has been studied as well, by postulating
oscillating control parameters (external fields) and considering the synchronization properties of the associated order parameters, including the Ising \cite{zhang2016critical,buendia2008dynamic}, the Blume--Capel \cite{keskin2005dynamic} and the Potts models \cite{mendes1991dynamics} and the Ginzburg--Landau theory \cite{robb2014extended}.

Alternatively, in the very seminal works on this topic the emergent character of oscillations (thus, called self-oscillations) without the need to postulate them was remarked \cite{andronov1966theory,strogatz2018nonlinear,Jenkins2013}. This area has been considerably developed during recent years and appearance and synchronization of periodic, quasiperiodic, chaotic and even hyperchaotic self-sustained oscillations in mechanical, physical and biological models, described by dynamical systems, has been demonstrated and studied \cite{Holms,strogatz2018nonlinear,Pikovsky,Rossler2020,borisyuk1992bifurcation}. Moreover, the importance of bifurcations or their sequences leading to the onset of regular or chaotic self-oscillations has been demonstrated in numerous works (see, e.g. \cite{strogatz2018nonlinear,Kuznetsov,Garashchuk2019,Garashchuk2020,Garashchuk2021,Kazakov2023}).
Therefore, following this route, in this work we study scenarios of the emergence of collective oscillations in minimal spin lattice models \cite{baxter2016exactly} in presence of a feedback  between the control and the order parameter trying to enforce homeostasis in the symmetric phase.
We show that the feedback can generically put these systems out-of-equilibrium, in particular triggering coherent collective oscillations.
The static counterparts of these system have a well defined free energy landscape and an important question is to what extent the latter can provide insights into the actual dynamics. We will illustrate this mechanism in the Landau theory with feedback that gets mapped into  nonlinear Van der Pol type oscillators. Then we will explore it concretely in the Ising, Blume--Capel and Potts models.

We demonstrate the onset of collective periodic oscillations in the Ising model with feedback after the corresponding mean field dynamical system undergoes the Andronov--Hopf bifurcation. We also obtain that for slow feedback the most probable value of a cycle's amplitude follows the Onsager law. For the Blume--Capel model we show that there is the Bautin bifurcation in the mean field dynamical system that correspond to the tricritical point in its collective counterpart. The existence of such bifurcation naturally leads to the presence of multistability in both mean filed and full microscopic models, which is numerically illustrated.
As far as the Potts model with three colors in concerned, we observe even more complex behaviour. For the regions with low feedback and high temperature we obtain that there is the onset of self-oscillations in the way similar to the Blume--Capel model and again with a region of multistability in the parameters' space. However, for bigger feedback we demonstrate that the behaviour is much more complex and quasiperiodic and chaotic oscillations emerge in both mean filed and microscopic models. Two bifurcation scenarios for the onset of chaotic oscillations are observed: the Fiegenbaum cascade of period doubling and the Aifraimovich--Shilnikov scenario of torus destruction. These results manifest that limit cycles bifurcations, complex bifurcation scenarios and collective quasiperiodic and chaotic oscillations can be observed in full microscopic models.
Furthermore, we demonstrate that the low-dimensional dynamical systems obtained via linear response provide for an effective mean field description of the complex behavior of the full system.
On the whole, we believe that this work provides a way of understanding the onset of the  complex collective behaviour in out-of-equilibrium  systems.

%We believe that this is the first time when it is shown that limit cycles bifurcations, complex bifurcation scenarios and collective chaotic oscillations can be observed in full microscopic models and their correspondence to the mean field equations is demonstrated.

%We believe that the results of this work establish a strong connection between bifurcations in dynamical systems and the onset of the collective behaviour in out-of-equilibrium spin systems.

The rest of the article is organized as follows. In the next Section we present our main results on collective oscillations in feedback lattice models. In the first Subsection we briefly discuss the feedback Landau theory. Subsection 2.2 is devoted to Ising model where we deal with the dynamics of the system in a fully connected geometry with a general linear feedback and in a 2D square lattice. The subsequent section 2.3 presents our results for the Blume-Capel model, that is an extension of the Ising model in presence of vacancies. Section 2.4 is focused on the Potts model with $q=3$ colors with feedback and  we summarize and discuss our findings in the Conclusion.

\section{Results}
\subsection{Feedback Landau theory}
Here we generalize the results presented in \cite{de2019feedback} on the effect of the presence of feedback in the Landau theory with higher order relevant terms. Let us consider the expansion of the free energy density up to the 6th power of a scalar order parameter $\phi$
\begin{equation}
\mathcal{L}(\phi) = -h\phi - \frac{\beta-1}{2} \phi^2 + \frac{a}{4} \phi^4 + \frac{b}{6} \phi^6,
\end{equation}
where $h$ is the external field, $\beta$ is the inverse temperature and $a,b$ are arbitrary real parameters.

Upon considering a negative feedback between $h$ and $\phi$ aiming at controlling the system into the symmetric phase, by linear response we have, upon neglecting noise in the thermodynamic limit, the dynamical system
\begin{align}
\dot{\phi} &= -\frac{\partial \mathcal{L}}{\partial \phi} = h + (\beta-1)\phi -a\phi^3 -b\phi^5, \nonumber \\
\dot{h} &= -c\phi.
\end{align}
The system has an equilibrium state at $(\phi=0,h=0)$ whose stability is associated to the eigenvalues of the Jacobian of the linearized system
\begin{equation}
\lambda_{\pm} = \frac{\beta-1 \pm \sqrt{(\beta-1)^2-4c}}{2}
\end{equation}
so, it is stable iff $\beta<1$, being a stable focus if $4c>(\beta-1)^2$ and a stable node otherwise. Since the system is confined (as it can be seen by looking to the gradient at the boundary of the square with vertex ), at $\beta_c=1$ the eigenvalues are purely imaginary and the real part changes sign and the system is undergoing the Andronov--Hopf bifurcation with emergent non-linear oscillations.

The character of the bifurcation can be assessed by calculating the first Lyapunov coefficient (see, e.g. \cite{Holms,Kuznetsov}), that in this case it can be calculated as
$l_1=-\frac{3}{8}a$. Thus, we have at $a=0$ the bifurcation of the system changes from supercritical to subcritical behavior with a discontinuous onset of oscillations.
This analysis seems to support an equivalence between the character of the Hopf bifurcations and the one of the underlying equilibrium static transition.  If that would be the case the supercritical emergence corresponds to a second order phase transition and subcritical emergence corresponds to a first order phase transition, where by the n-th order we mean the order of singularity of the free energy function, in particular, independently of the feedback strength $c$. We will show that this Landau feedback theory captures the qualitative behavior of the Ising and Blume--Capel models while it fails to do so for the Potts model as soon as $q=3$.

\subsection{Feedback Ising model}

We will extend in this section the results of \cite{de2019feedback} analytically in presence of a general linear control and numerically in finite dimension.
\begin{figure*}[!ht]\label{fig1}
\includegraphics*[width=0.9\textwidth]{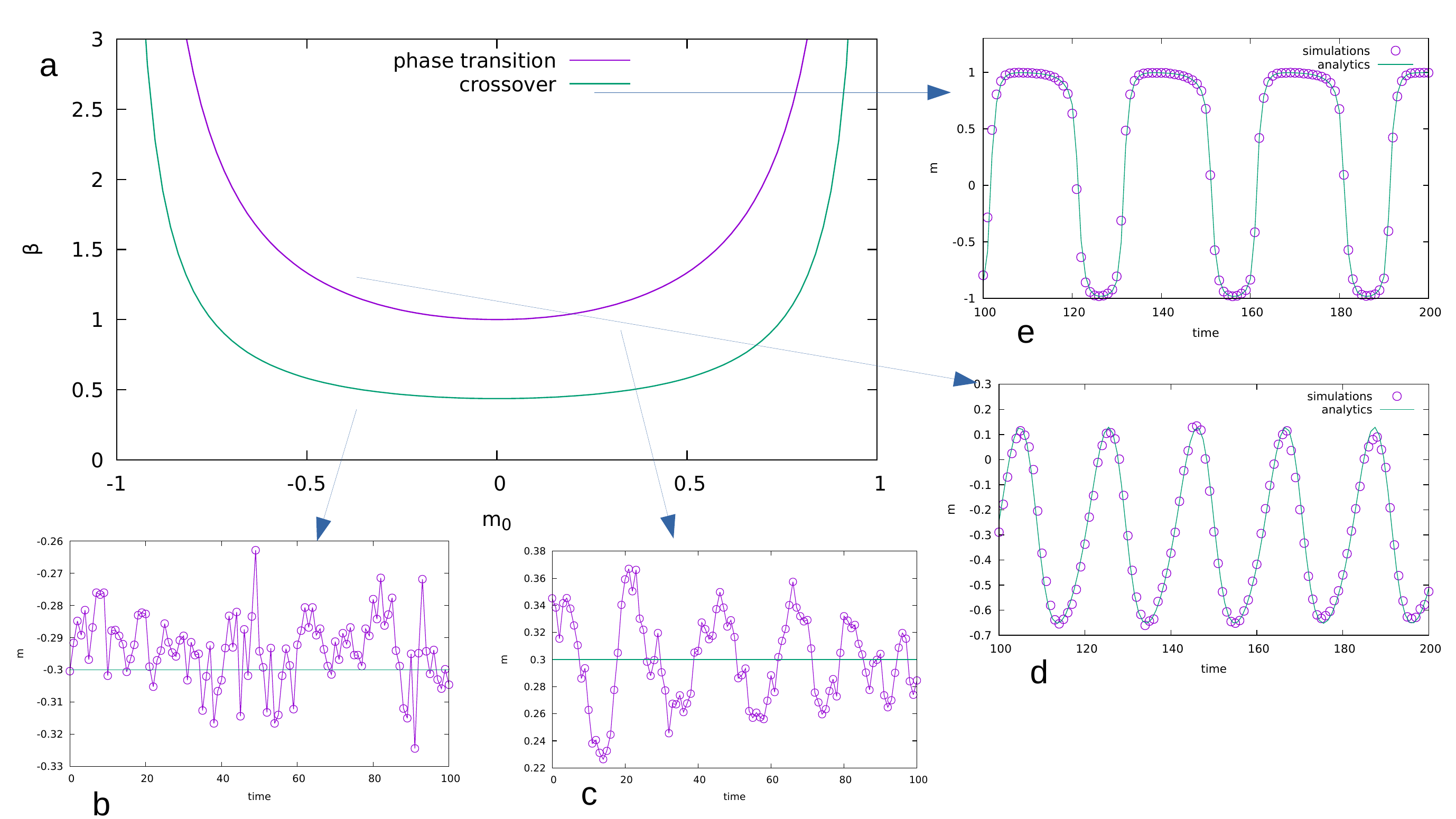}
\caption{(a): Mean field phase diagram of the feedback Ising model in the plane $(m_0,\beta)$ for $J=1$, $c=0.1$, both the critical line and the dynamical crossover line are highlighted. (b,c,d,e): magnetization time traces of the system ($N=10^4$ spins) simulated on a fully connected geometry in four different points corresponding to different dynamical behavior. (b): $m_0=-0.3$, $\beta=0.25$. Simulations are made via the Metropolis--Hastings method \cite{metropolis1953equation,hastings1970monte}}
\end{figure*}

We consider the Ising model, i.e. a model of $Z_2$ spins $s_i=\pm1$ variables sitting on the nodes of a lattice graph whose Hamiltonian is
\begin{equation}
H = -J\sum_{\langle i,j \rangle} s_i s_j -h \sum_i s_i,
\end{equation}
where the bracket in the first sum stands for the graph bonds and we indicate with $J$ and $h$ the interaction strength and the external magnetic field, respectively. Upon generalizing from \cite{de2019feedback} we apply a negative feedback between the instantaneous magnetization $m = \frac{1}{N} \sum_i s_i$ and the external magnetic field $h$, with the aim of setting the former to a prescribed value $m_0$, $|m_{0}|<1$, that is a parameter of the system.

We consider the mean-field Curie--Weiss approximation, where in the limit of large $N$ the free energy of the system is
\begin{equation}
f = -\frac{J}{2} m^2 +\frac{1}{\beta}\log(\cosh(\beta(Jm+h))).
\end{equation}
This is equivalent to consider the system in a fully connected graph and rescaling the interaction energy $J\to J/N$.
The free energy  is obtained from the partition function
 \cite{baxter2016exactly}
\begin{equation}
Z = \sum_{\{\vec{s}\} } e^{-\beta H} = \int dm e^{-N\beta f(m)}
\end{equation}
In regard to the dynamics we will assume linear response, i.e. that the time derivative of the magnetization is proportional to the gradient of the free energy function \cite{zwanzig2001nonequilibrium}. Upon considering the feedback the system is described by  the equations
\begin{eqnarray}
\label{eq:Ising_f}
\dot{m} = -m + \tanh (\beta(Jm+h)), \nonumber \\
\dot{h} = -c(m-m_0).
\end{eqnarray}
This system admits the only stationary point
\begin{eqnarray}
m_s = m_0, \quad h_s = \mbox{atanh}(m_0)/\beta -Jm_0 ,
\end{eqnarray}
whose linear stability can be assessed studying the eigenvalues of the Jacobian matrix ($\beta_2 = \beta J (1-m_0^2)$)
\begin{equation}
\lambda_\pm = \frac{\beta_2-1 \pm \sqrt{(\beta_2-1)^2 -4\beta_2 c}}{2}
\end{equation}
The equilibrium point is  stable iff $\beta<\beta_c = \frac{1}{J(1-m_0^2)}$ and its character changes from a node to a focus (dynamical crossover), where eigenvalues develop an imaginary part, if $c>\frac{(\beta/\beta_c-1)^2}{4 \beta/\beta_c}J$. In the latter case the loss of stability at $\beta=\beta_c$ (phase transition) implies an Andronov--Hopf bifurcation triggering self oscillations.
The character of the bifurcation, supercritical (continuous) or subcritical (discontinuous) can be assessed from the sign of the first Lyapunov coefficient $l_1$. For \eqref{eq:Ising_f} at $\beta=\beta_{c}$ it is
\begin{equation}
    l_{1}=-\frac{c+J}{\left(cJ\right)^{3/2}(1-m_{0}^{2})}.
\end{equation}
Consequently, we see that the Andoronov--Hopf bifurcation is supercritical provided that $|m_{0}|<1$.

% Upon following  we define  $\omega = \sqrt{c/J}$ and
% \begin{equation}
% f(z) = -z + (1-m_0^2) \frac{\tanh{\beta J z}}{1+m_0 \tanh{\beta J z}}
% \end{equation}
% from which it can be easily shown that
% the first Lyapunov coefficient is given by the formula
% \begin{equation}
% \mathsf{l_1} = \frac{1}{16}(1+\omega^2) (f''' -(f'')^2)
% \end{equation}
% calculated at the equilibrium point at the bifurcation  $z=0,\beta=\beta_c$.

Beyond the critical point, an approximate analytical solution can be worked out for $\beta>\sim\beta_c$ (harmonic oscillations) by a two-time expansion \cite{strogatz2018nonlinear}, if we call $\epsilon = \beta-\beta_c$ we have
\begin{equation}
m-m_0 \sim \sqrt{\epsilon} \cos{((1+\frac{1}{2}\epsilon)\sqrt{c}t + \phi_0)}
\end{equation}
On the other hand, in the limit $\beta \to \infty$, where the system is performing relaxational oscillations, the  equations are piece-wise linear and the shape of the limit cycle can be found by matching the boundary conditions of  solutions in half-spaces \cite{andronov1966theory}.
Notice that the two regimes differ qualitatively by the fact that the quasi-harmonic
oscillations are centered in $m_0$ and pass equal time up and down from this value,
while in the relaxational regime they are centered in $0$ and pass uneven times upon
having positive and negative values such that their average will be in the end $m_0$.

\begin{figure*}[!ht]\label{fig2}
\begin{center}
\includegraphics*[width=0.9\textwidth,angle=0]{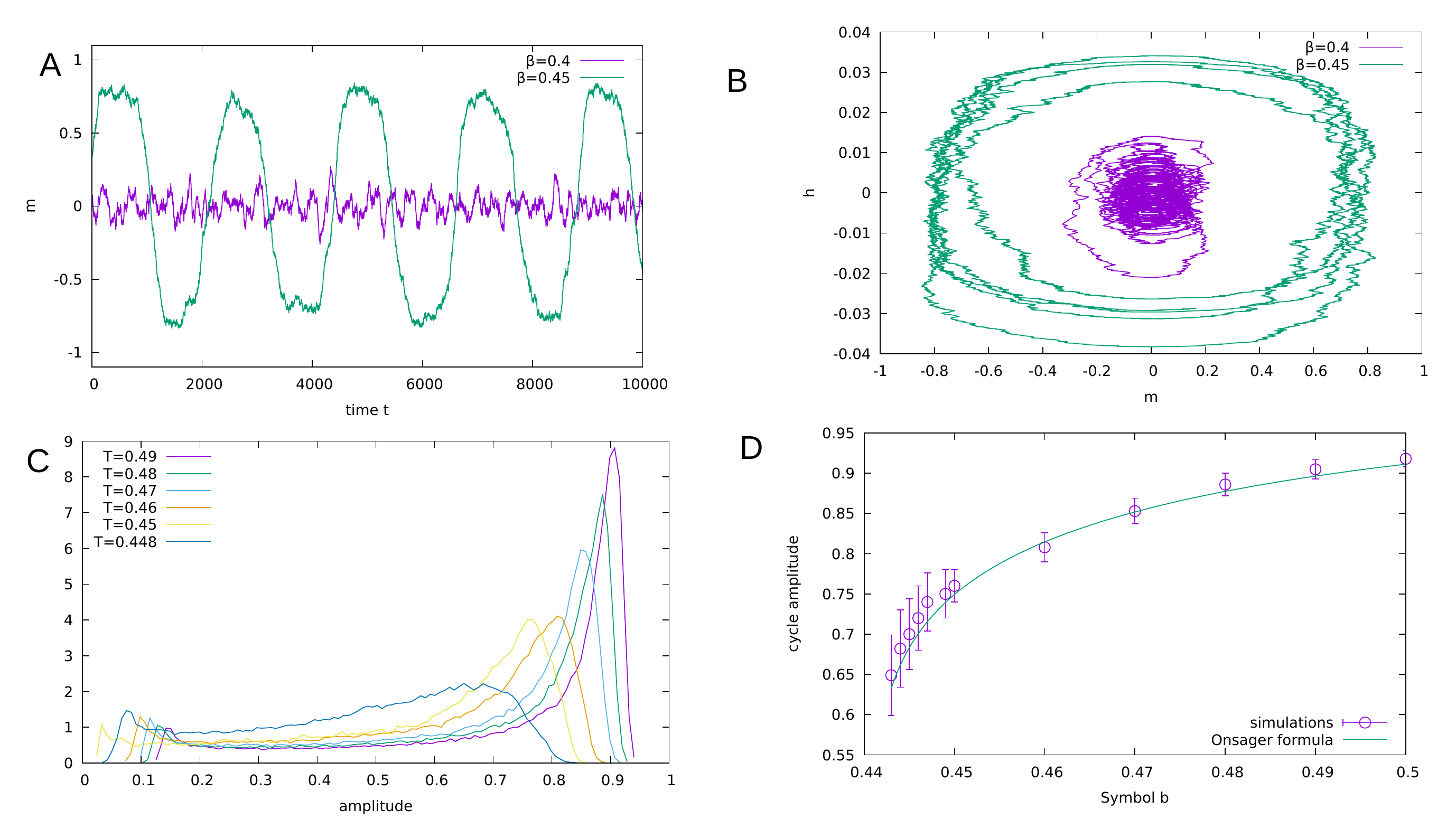}
\caption{Simulations of the feedback Ising model on a 2D square lattice; system size $N=10^{4}$ spins, feedback strength $c=10^{-4}$ (A): Magnetization as function of time $m(t)$ for $\beta=0.47,0.45$. (B) Trajectories in the phase plane $(m,h)$ for  $\beta=0.47,0.45$. (C) histograms of the probability distribution of the limit cycle amplitude at several $\beta$s. (D) Most probable limit cycle amplitude  as function of $\beta-\beta_c$ compared with Onsager formula and $\beta_c=\frac{\log(1+\sqrt{2})}{2}$. Simulations are made via the metropolis-hastings method \cite{metropolis1953equation,hastings1970monte}.}
\label{fig:fig2}
\end{center}
\end{figure*}

These results are in quantitative agreement with numerical simulations on the lattice system in a fully connected geometry as we show in  Fig. 1 where we recapitulate the behavior in a phase diagram in the plane $(m_0,\beta)$.
The mean-field approximation of the static Ising model is known to capture qualitatively the behavior of the system  in finite dimensional geometry, but it
fails  quantitatively on the surroundings of the critical point.
The study of the Ising model in finite dimension is arguably one of the most important areas  in statistical  physics, touching upon issues related to  field theories and   renormalization . One of the most important and earlier results, due to Onsager \cite{onsager1944crystal} and achieved by combinatorial counting, is an analytical solution for the  system in a 2D square lattice where we have the formula for the spontaneous magnetization:
\begin{equation}
m  =     (1-\sinh^{-4}{2\beta})^{1/8},
\end{equation}
valid for
\begin{equation}
\beta>\beta_c= \frac{1}{2} \log (1+\sqrt{2})
\end{equation}

The analogy with the static counterpart would suggest that for the case with feedback in finite dimension we could have a limit cycle emerging with a non-trivial exponent and we  explored this issue by numerically studying the  Ising model with feedback in a 2D square lattice graph.

The corresponding results are shown in Fig. \ref{fig:fig2}. Simulations have been performed for a system with $N=10^4$ spins, for a neutral  $m_0=0$ and slow $c=10^{-4}$ feedback. First of all we provide evidence (see fig 2A and 2B) that for  $\beta>\beta_c$ a limit cycle emerges whereas for $\beta<\beta_c$ the dynamics is relaxing into fixed point with $\beta_c\simeq 0.44$ consistent with the analytical value.

We investigate in further detail this aspect by quantitatively comparing the limit cycle amplitude with the Onsager formula. This has been done by observing the trajectory of the system in the phase space $(m,h)$ and  extracting the radial coordinate. The distribution of the latter is shown in fig 2C and for $\beta>\beta_c$ it develops a second peak in correspondence with the formation of a limit cycle. The value at peak has been evaluated against $\beta$ in fig 2D and the trend is compared with Onsager formula: we find a good agreement within errors, evaluated as the width at half maximum.

\subsection{Feedback Blume--Capel model}
Here we consider a generalization of the Ising model in presence of vacancies, i.e. the spin variables admit the null value $s_i=0,\pm1$ and the Hamiltonian has an additional term counting the number of filled sites whose average is fixed by the chemical potential $\Delta$:
\begin{equation}
H = -J\sum_{\langle i,j \rangle} s_i s_j -h \sum_i s_i + \Delta \sum_i s_i^2
\end{equation}

The static free energy of the system in a fully connected geometry can be calculated analytically \cite{blume1966theory}
\begin{equation}
f = -\frac{J}{2} m^2 +\frac{1}{\beta}\log(1+2\cosh(\beta(Jm+h))e^{\beta \Delta})
\end{equation}
The system with feedback fulfills the approximate equations
\begin{eqnarray}
\label{eq:BCm}
\dot{m} &=& -m + \frac{\sinh(\beta(Jm+h))}{\frac{e^{\beta \Delta}}{2}+\cosh(\beta(Jm+h))}, \\
\dot{h} &=& -cm. \nonumber
\end{eqnarray}
%The equilibrium point $m_s=0,h_s=0$ looses its stability on the critical line
%\begin{equation}
%\Delta_c = \log(2(\beta-1))/\beta
%\end{equation}

In order to simplify analysis of \eqref{eq:BCm} we introduce new variables $m'=\beta J m$ and $h'=\beta h$. We also redefine the parameters as follows: $c'=c/J>0$, $\beta'=J\beta>0$, $\Delta'=\Delta/J$ and ${\rm e}^{\beta'\Delta'}=2\nu>0$. As a result, from \eqref{eq:BCm} we obtain (the primes are omitted)
\begin{equation}
\begin{gathered}
\label{eq:BCm_non_dim}
\dot{m} = -m + \frac{ \beta \sinh(m+h)}{\nu+\cosh(m+h)}, \\
\dot{h} = -cm.
\end{gathered}
\end{equation}
It is clear that the origin $O=(0,0)$ is the fixed point of \eqref{eq:BCm_non_dim}. We begin with the analysis of co-dimension one bifurcation in \eqref{eq:BCm_non_dim} and suppose that the $\beta$ is the control parameter and $c$ and $\nu$ have some fixed values. The Jacobi matrix for \eqref{eq:BCm_non_dim} at $O$ is
\begin{equation}\label{eq:J}
\begin{gathered}
  \mathcal{J} = \left(
  \begin{array}{cc}
              \frac{\beta-\nu-1}{\nu+1} & \frac{\beta}{\nu+1}\\
              -c& 0
            \end{array}\right).
 \end{gathered}
\end{equation}
Consequently, the eigenvalues of \eqref{eq:J} can be expressed via
\begin{equation}
    \sigma={\rm tr}\left(\mathcal{J}\right)=\frac{\beta-\nu-1}{\nu+1}, \quad
    \delta=\det \mathcal{J}=\frac{\beta c}{\nu+1},
\end{equation}
as follows
\begin{equation}
    \lambda_{1,2}=\frac{1}{2}\left(\sigma\pm\sqrt{\sigma^{2}-4\delta}\right).
\end{equation}

\begin{figure*}[!ht]
    \centering
    \includegraphics[width=0.9\textwidth]{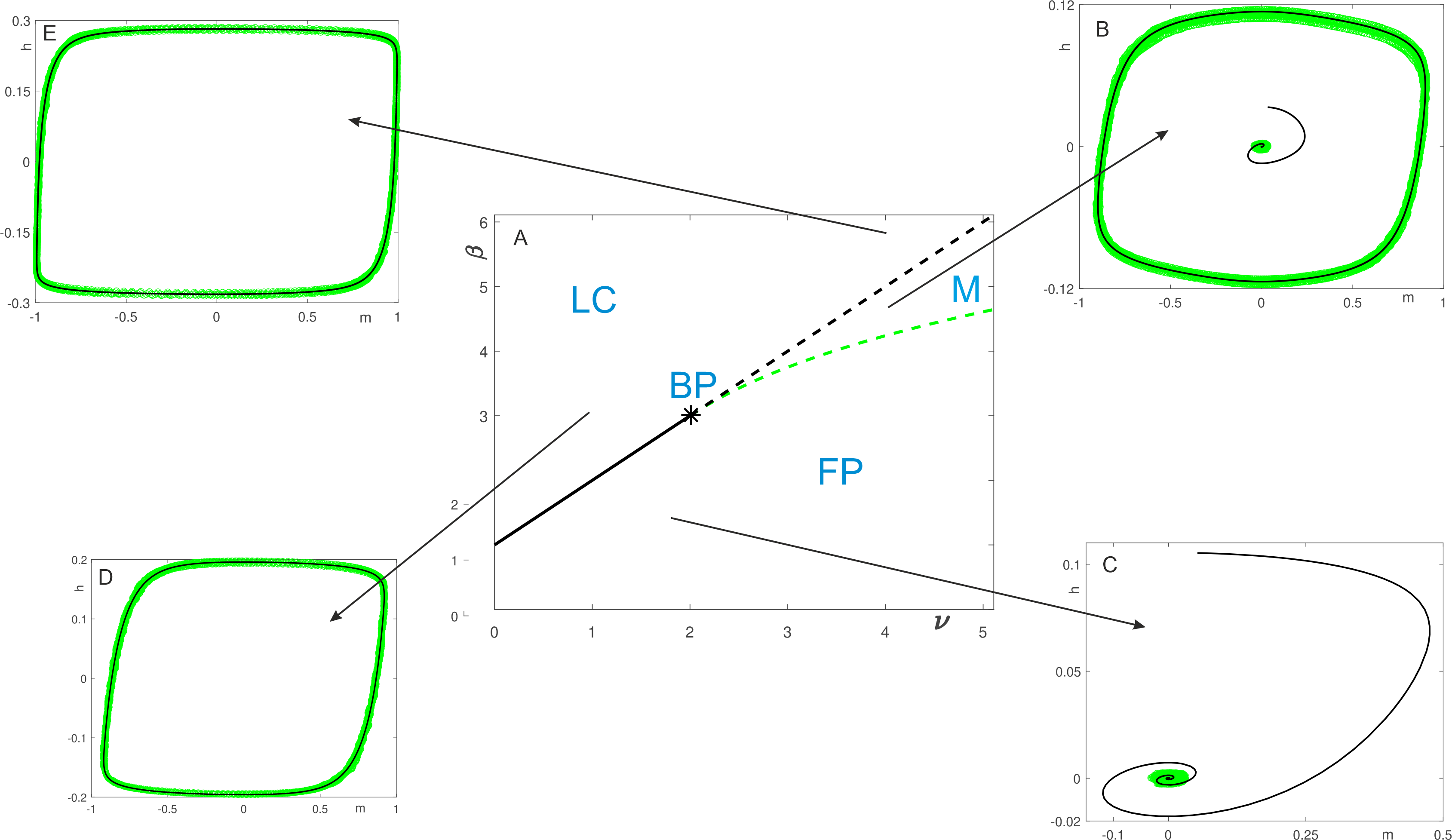}
    \caption{Sketch of the behaviour of system \eqref{eq:BCm_non_dim} nearby the Bautin bifurcation point: A - bifurcation diagram for \eqref{eq:BCm_non_dim}, where black line is the line of the Andoronov--Hopf bifurcation and the green one is the line of saddle-node bifurcation of limit cycles. Phase portraits of solutions of \eqref{eq:BCm_non_dim} nearby bifurcation point (black line) and the results full microscopic stochastic simulations in the fully connected model with $N=10^4$ spins (green dots): B -confirmation of multistability; C -stable fixed point; D, E - different limit cycles. Simulations are made via Metropolis--Hastings method \cite{metropolis1953equation,hastings1970monte}}
    \label{fig:fig3}
\end{figure*}

First, we are interested in the Andronov--Hopf bifurcation and, hence, we assume that $4\delta-\sigma^{2}>0$, so that we have complex conjugate eigenvalues. These means that the $\beta$ lies in the interval
\begin{equation}
    (2c+1-2\sqrt{c(c+1)})(\nu+1)<\beta< (2c+1+2\sqrt{c(c+1)})(\nu+1).
\end{equation}
The first condition for the Andronov--Hopf bifurcation $\sigma(\beta_{0})=0$ results in $\beta_{0}=\nu+1$. The second condition $\delta(\beta_{0})=c>0$ holds automatically. The non-degeneracy conditions $\mu'(\beta_{0})=\sigma'(\beta_{0})/2\neq0$ and $l_{1}(\beta_{0})\neq0$ hold at $\nu\neq2$. Indeed, $2\mu'(\beta_{0})=1/(\nu+1)>0$ and the first Lyapunov coefficient for \eqref{eq:BCm_non_dim} is given by
\begin{equation}
    \label{eq:fist_L_coeff}
    l_{1}(\beta_{0})=\frac{(\nu-2)(c+1)}{4c^{\frac{3}{2}}(\nu+1)}\neq 0, \quad \nu\neq2.
\end{equation}
We see that $l_{1}\neq0$ except at $\nu=2$, when the Andronov--Hopf bifurcation switches from supercritial to subcritical one.

At $\nu=2$ the first Lyapunov coefficient vanished and we have that $\beta=3$ and $\nu=2$ is the point of the Bautin bifurcation (see, e.g. \cite{Kuznetsov}). In order to check the non-degeneracy conditions at the point of the Bautin bifurcation we compute the second Lyapunov coefficient for \eqref{eq:BCm_non_dim} at $\beta=3$ and $\nu=2$, which is $l_{2}=-\sqrt{c}(c+1)^{2}/18<0$  for $c>0$. Consequently, the line $\beta=\nu+1$ is the line of the Andronov--Hopf bifurcation with the Bautin point at $(2,3)$ that separates the supercritical part of the line from the subcritical one.

In Fig. \ref{fig:fig3} we demonstrate bifurcation diagram for \eqref{eq:BCm_non_dim} at $c=0.01$. The black line is the line of the Andronov--Hopf bifurcation, where continuous part corresponds to supercritical bifurcation and broken part to subcritical one. The green line is the line of saddle-node bifurcation of limit cycles, which is computed numerically with the help of the MATCONT \cite{Matcont}. The star denotes the Bautin point. One can see that parameters space is separated into three regions. Broken black and green lines form a region of multistability, where a stable limit cycle coexists with a stable fixed point. Below solid black and green broken lines the dynamics is defined by a stable fixed point. Above the black line there exists a limit cycle, the correspond to oscillations in both \eqref{eq:BCm_non_dim} and full stochastic microscopic model.

Finally, let us remark that the Bautin bifurcation for a spin system has been also described in \cite{collet2019effects}, where they consider a dissipative term in the Curie--Weiss model in presence of local random fields.

\subsection{Feedback Potts model}

In the Potts model the lattice variables can assume one of $q$ given states (``colors'') $s_i=1 \dots q$. Variables with equal colors in neighbouring sites lower the energy by a factor $J$ and we consider $q$ external fields $h_a$
fixing the relative color fractions $x_a=\frac{1}{N} \sum_i \delta_{s_i,a}$  in the system. The corresponding the Hamiltonian is
\begin{equation}
H = -J \sum_{\langle i,j\rangle } \delta_{s_i,s_j} -\sum_a h_a \sum_i \delta_{s_i,a}.
\label{hamipott}
\end{equation}
In the fully connected case the expression for the free energy  is \cite{wu1982potts}:
\begin{equation}
A =  -\frac{J}{2}\sum_a x_a^2 -\sum_a h_\sigma x_a + T \sum_a x_a \log x_a +\lambda \left(\sum_a x_a -1\right)
\end{equation}
where $x_a$ are the fractions of spins belonging to the same colour $a$ and the last term is a Lagrange multiplier enforcing normalization.
Following the same scheme used for the Ising model we have that the dynamics shall approximately follow the set of equations
\begin{equation}
\begin{aligned}
\dot x_a &= -x_a + \frac{e^{\beta(J x_a+h_a)}}{W},  \\
\dot h_a &= -c(x_a - 1/q),\hspace{5mm} a = 1,\dots,q, \\
W &= \sum_a e^{\beta(J x_a+h_a)}.
\end{aligned}
\label{eqmotonaive}
\end{equation}
which describes the motion for a linear response approximation.

 \begin{figure*}[!ht]
    \centering
     \includegraphics[width=0.9\textwidth]{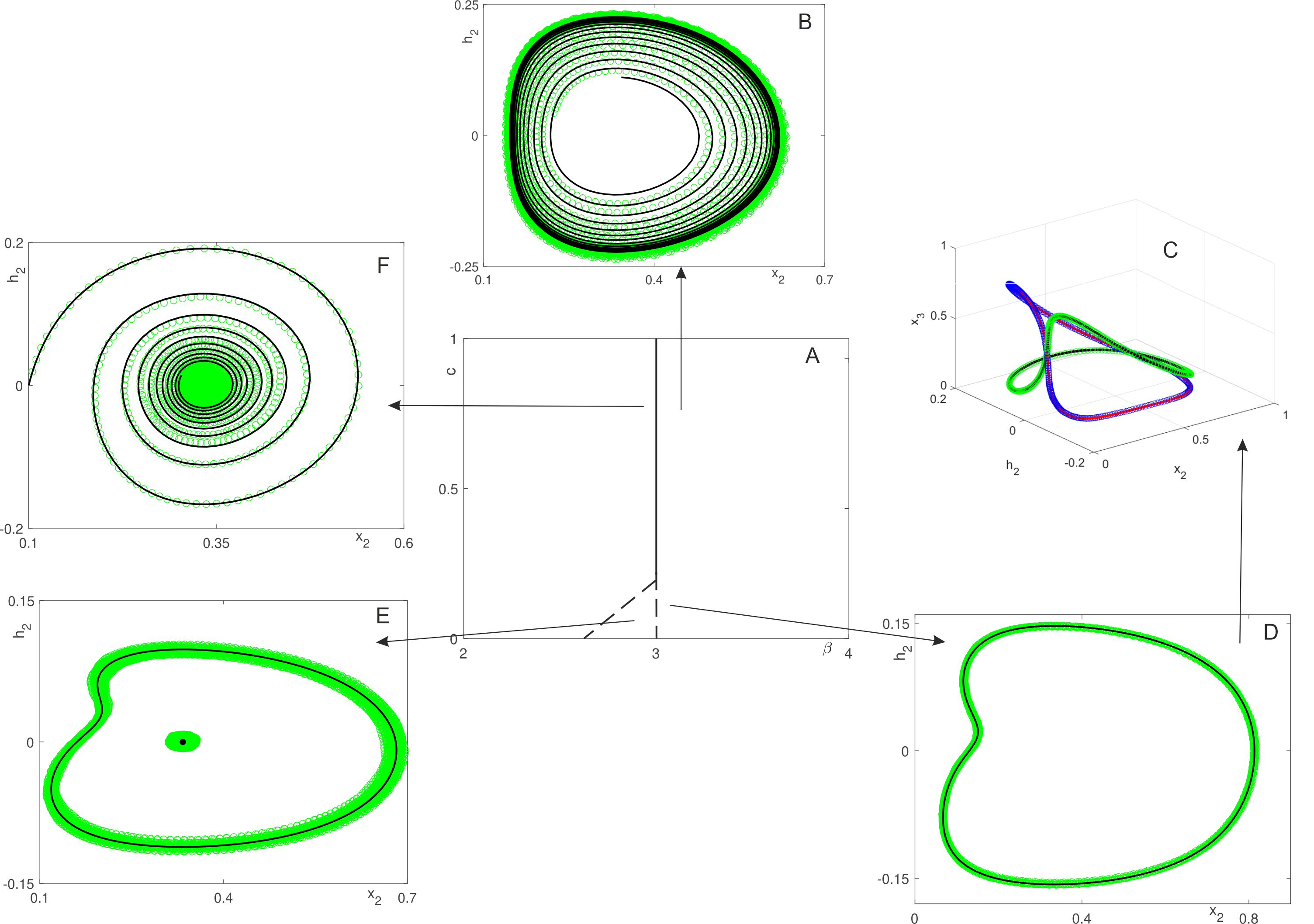}
     \caption{The behaviour of system \eqref{eq:Potts3_1} in the vicinity of the bifurcation line $\beta=3$. A: a sketch of a bifurcation diagram illustrating transition from a subcritical to supercritical bifurcation; B-E phase portraits of some attractors in the vicinity of  $\beta=3$. Black lines represent numerical solutions of \eqref{eq:Potts3_1}, while green dots are obtained from fully microscopic stochastic simulations of a system of size $N=10^5$ spins via the Metropolis--Hastings method.}
     \label{fig:fig4a}
 \end{figure*}

System \eqref{eqmotonaive} possesses two conservation law, that are
\begin{equation}
    \label{eq:Potts_c1}
    \sum_{a}x_{a}=1+C_{0}{\rm e}^{-t},
\end{equation}
and
\begin{equation}
    \label{eq:Potts_c2}
    \sum_{a}h_{a}=C_{1},
\end{equation}
where $C_{0}$ and $C_{1}$ are arbitrary constants.

 \begin{figure*}[!ht]
    \centering
     \includegraphics[width=0.9\textwidth]{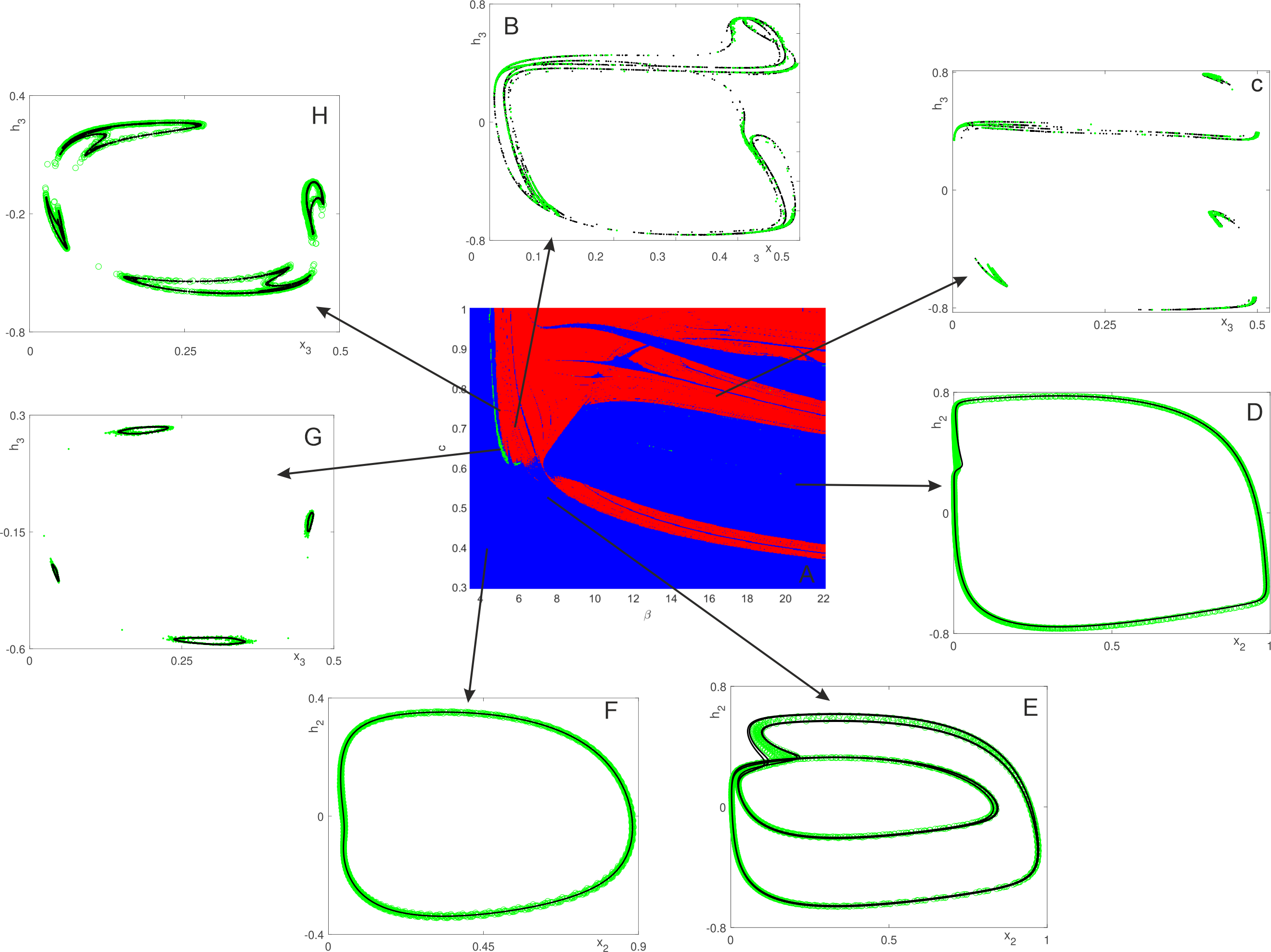}
     \caption{Plate A represents the two-dimensional chart of the Lyapunov exponents for \eqref{eq:Potts3_1}, Blue color corresponds to the periodic dynamics, green to quasyperiodic and red to chaotic one.  Plates D,E,F correspond to the phase portraits of periodic attractors at
     $\beta=4$, $c=0.4$, $\beta=8.1$, $c=0.55$, $\beta=21$, $c=0.6$, respectively. In plates B,C,G,H we present Poincar\'e sections of three chaotic and one quasiperiodic attractors at $\beta=9.1$, $c=0.9$, $\beta=20$, $c=0.4$, $\beta=5.2$, $c=0.65$, $\beta=5.1$, $c=0.75$. Black lines corresponds to numerical solutions of system \eqref{eq:Potts3_1} and green dots are obtained from fully microscopic stochastic simulations of a system of size $N=10^6$ spins via the Metropolis-Hastings method.}
     \label{fig:fig4}
 \end{figure*}

Since at $t=0$ the sum of all $x_{a}$ is equal to $1$, we set $C_{0}=0$. Moreover, transformation $h_{a}\rightarrow h_{a}+\mbox{const}$ results in additional constant term in Hamiltonian \eqref{hamipott} and, hence, does not affect the dynamics. Consequently, without loss of generality, we assume that $C_{1}=0$ in \eqref{eq:Potts_c2}.

Now we consider the case of three colors, i.e. $q=3$. Taking into account \eqref{eq:Potts_c1} and \eqref{eq:Potts_c2} and rescaling variables as follows $h_{a}=Jh_{a}^{'}$, $\beta J=\beta^{'}$, $c/J=c^{'}$ from \eqref{eqmotonaive} at $q=3$ we obtain (the primes are omitted)
\begin{eqnarray}
\label{eq:Potts3_1}
\dot{x}_{2} &=& -x_{2}+\frac{{\rm e}^{\beta(x_{2}+h_{2})}}{\widetilde{W}},\\
\dot{h}_{2} &=& -c \left(x_{2}-\frac{1}{3}\right),\\
\dot{x}_{3} &=& -x_{3}+\frac{{\rm e}^{\beta(x_{3}+h_{3})}}{\widetilde{W}},\\
\dot{h}_{3} &=& -c \left(x_{3}-\frac{1}{3}\right),
\end{eqnarray}
and
\begin{equation}
\widetilde{W}={\rm e}^{\beta(1-x_{2}-x_{3}-h_{2}-h_{3})}+{\rm e}^{\beta(x_{2}+h_{2})}+{\rm e}^{\beta(x_{3}+h_{3})}
\end{equation}
System \eqref{eq:Potts3_1} is symmetric with respect to swapping of indices ($2\longleftrightarrow3$) and has one equilibrium point $A=(1/3,0,1/3,0)$. The eigenvalues of the Jacobi matrix at this fixed point are
\begin{equation}
    \lambda_{1,2,3,4}=\frac{\beta-3}{2}\pm\frac{\sqrt{\beta^{2}-6(2c+1)\beta+9}}{6}.
\end{equation}
Suppose that
\begin{equation}
\beta\in\left(3[2c+1-2\sqrt{c^{2}+c}],3[2c+1+2\sqrt{c^{2}+c}]\right)   \end{equation}
then the eigenvalues are complex conjugated. Passing through $\beta=3$ the real part of eigenvalues cross the imaginary axis and, hence, the fixed point $A$ losses its stability. Due to the symmetry of \eqref{eq:Potts3_1}, the Jacobi matrix at $A$ has multiple eigenvalues and we have the resonant double Hopf bifurcation (see, e.g. \cite{Broer2021} and references therein). Consequently, the analytical treatment of the behaviour of \eqref{eq:Potts3_1} near $\beta=3$ is connected with some difficulties. However, numerically we observe that the dynamics in the vicinity of the bifurcation line $\beta=3$ is similar to those of the Blume--Capel system near the line of the Andoronov--Hopf bifurcation (see, Fig.  \ref{fig:fig4a} and cf. Fig. \ref{fig:fig3}).
From Fig.\ref{fig:fig4a} we see that in the lower left part of the bifurcation line $\beta=3$ there is a region of multistability  (see, Fig.  \ref{fig:fig4a}E, D), where a fixed point coexists with a periodic orbit. This suggest that there exists some $c^{*}$ such that for $c<c^{*}$ the bifurcation at $\beta=3$ is subcritical. Increasing the value of $c$ we observe that the bifurcation type switches from subcritical to supercritical one: there is no multistaability  and at $\beta=3$ a stable small-amplitude limit cycle is born (see, Fig.  \ref{fig:fig4a} B, F). Let us also remark that due to the symmetry of \eqref{eq:Potts3_1} there are always two coexisiting orbits, which can be seen from Fig. \ref{fig:fig4a} plates C, D.

In order to study possible type of dynamics that appear in \eqref{eq:Potts3_1} away from the line $\beta=3$ we compute the dependence of the Lyapunov spectrum on the parameters $\beta$ and $c$ for the region $(\beta,c) \in [3,5,22]\times[0.3,1]$. For computation of the Lyapunov spectrum we use the standard algorithm by Bennetin et al. (see \cite{Benettin1980} ). We present the corresponding two dimensional chart of Lypaunov exponents in Fig. \ref{fig:fig4}. We color each point in this chart according to the signs of two largest Lyapunov exponents as follows:
\begin{itemize}
    \item $\lambda_{1}=0$, $0>\lambda_{2}>\lambda_{3}>\lambda_{4}$: periodic regime and blue color;
     \item $\lambda_{1}=\lambda_{2}=0 $, $0>\lambda_{3}>\lambda_{4}$: quasiperiodic regime and green color;
     \item $\lambda_{1}>0$, $\lambda_{2}=0$, $0>\lambda_{3}>\lambda_{4}$: chaotic regime and red color.
\end{itemize}

In Fig. \ref{fig:fig4} we also demonstrate phase portraits and Poincare sections of typical attractors that appear in \eqref{eq:Potts3_1}. Throughout this work the Poincar\'e map is constructed by considering intersections of the flow governed by \eqref{eq:Potts3_1} with the plane $x_{2}=1/2$, if it is not stated otherwise.

\begin{figure}[!ht]
    \centering
     \includegraphics[width=0.75\textwidth]{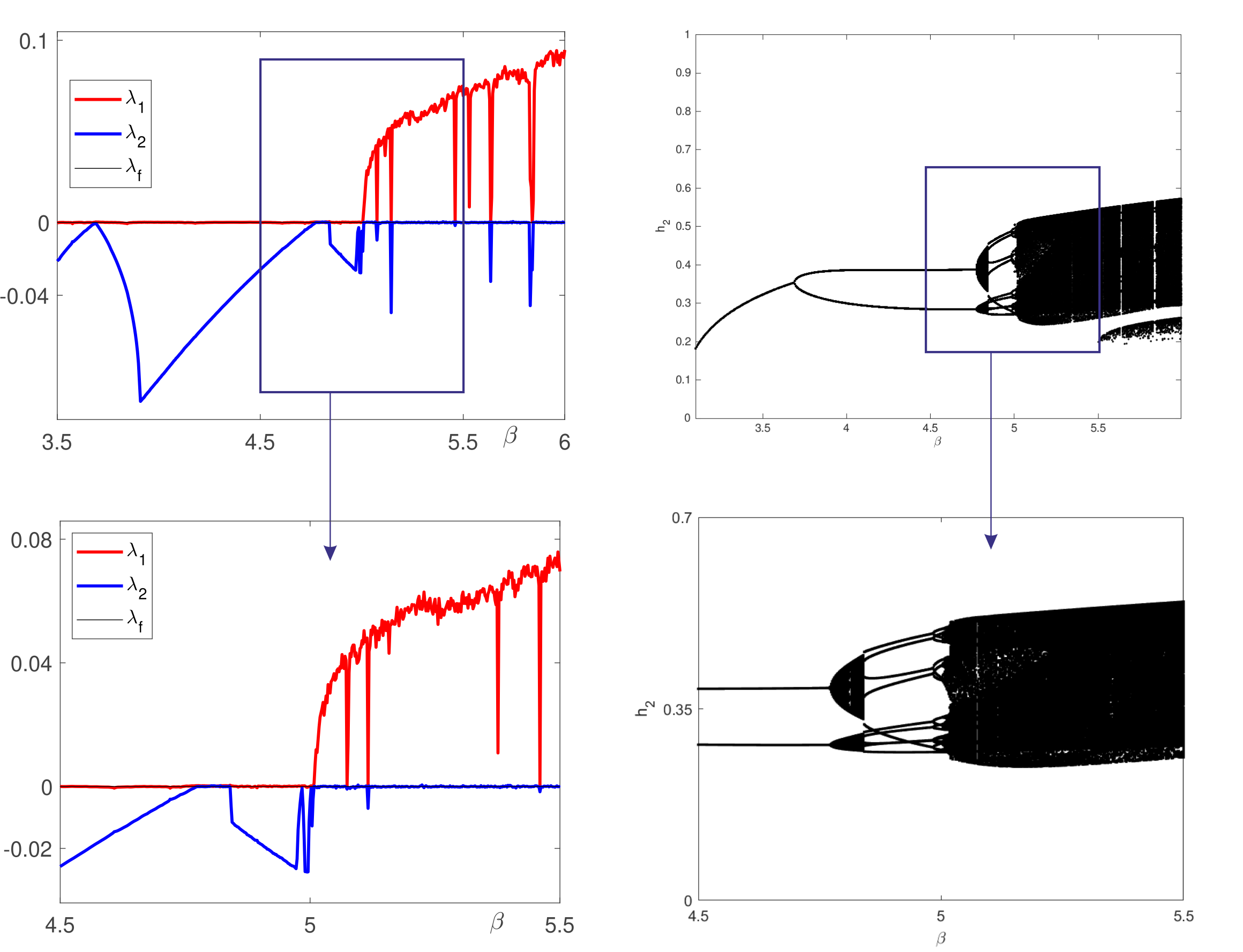}
     \caption{Lyapunov specta and bifurcation trees for system \eqref{eq:Potts3_1} at $c=0.75$ and $\beta\in[3.5,6]$.}
     \label{fig:fig5}
 \end{figure}

The vast blue regions in Fig. \ref{fig:fig4} correspond to the existence of a stable periodic orbit. There are also two separated red regions of the chaotic dynamics. The appearance of the chaotic attractors in these regions is governed by different scenarios. In the upper left region of  Fig. \ref{fig:fig4} we see that a thin green stripe of quasiperidic dynamics that is adjacent to the thin region of periodic behaviour with chaotic one next to it. This suggests that chaotic attractors can appear through the Afraimovich--Shilnikov scenario of am invariant torus destruction (see, e.g. \cite{Afraimovich1991,Garashchuk2019,Stankevich2021}).

\begin{figure*}[!ht]
    \centering
     \includegraphics[width=0.8\textwidth]{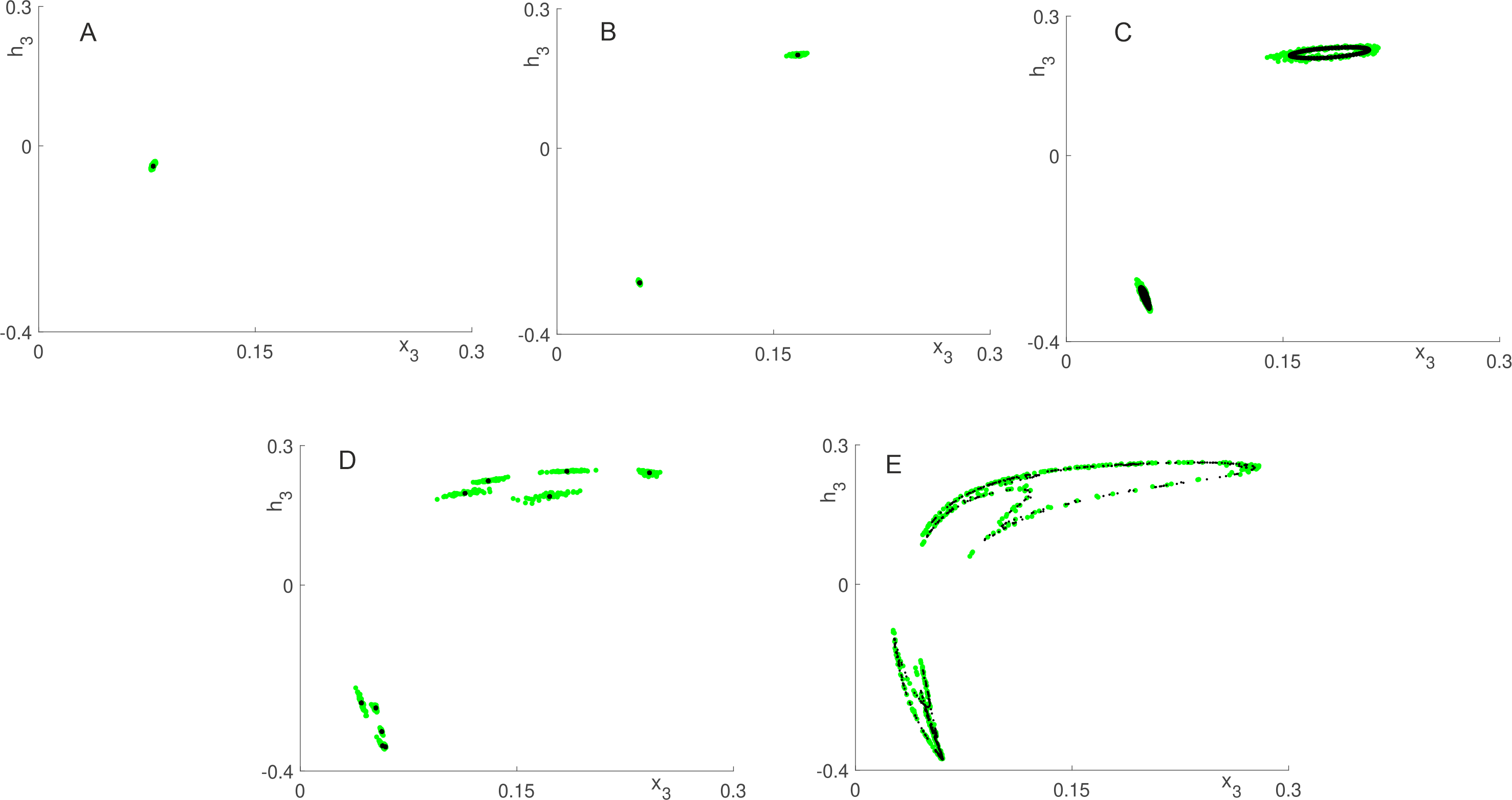}
     \caption{Poincare sections of the attactors that appear along the line of the Afraimovich--Shilnikov bifurcation scenario at $c=0.75$: A - a stable periodic orbit at $\beta=3.6$; B -a stable periodic orbit after period doubling at $\beta=4.5$; C -a stable quasiperiodic orbit at $\beta=4.8$; D  -a resonant periodic orbit at $\beta=4.9$; E - a chaotic attractor at $\beta=5.1$. By black lines  we show a numerical solutions of \eqref{eq:Potts3_1} and green dots are obtained from fully microscopic stochastic simulations of a system of size $N=10^6$ spins via the Metropolis-Hastings method.}
     \label{fig:fig6}
 \end{figure*}

\begin{figure}[!ht]
    \centering
     \includegraphics[width=0.7\textwidth]{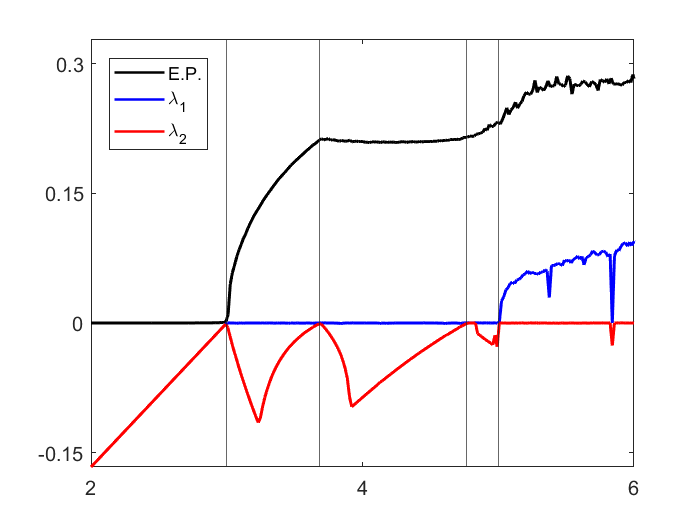}
     \caption{Entropy production for the Potts model and the Lyapunov spectrum for \eqref{eq:Potts3_1} at $c=0.75$ and $\beta\in[2,6]$.}
     \label{fig:fig7}
\end{figure}

In the region of lower feedback one can observe a big blue region of periodic oscillations next to a narrow red strip of chaotic ones. From Fig. \ref{fig:fig4}F,E one can see that if we approach this red region from the right the period of oscillations increases. Therefore, one can expect the appearance of the cascade of period doubling bifurcations, which we confirm below.

Let us begin with the onset of chaotic oscillations in the upper left part of the bifurcation diagram (see Fig. \ref{fig:fig4}). We see that inside the blue regions adjacent to red one there is a narrow green path. The left border of this green patch is the line of the supercritical Neimark--Sacker bifurcation, that corresponds to the birth of a stable quasiperiodic regime, which is called an invariant torus. The right border of the green region represents the formation of resonant stable and unstable periodic orbits that appear through the saddle-node bifurcation. Then, if we move further to the right in the  bifurcation diagram,
the stable resonant orbit becomes chaotic and so-called torus-chaos attractor appears.

 \begin{figure}[!ht]
    \centering
     \includegraphics[width=0.9\textwidth]{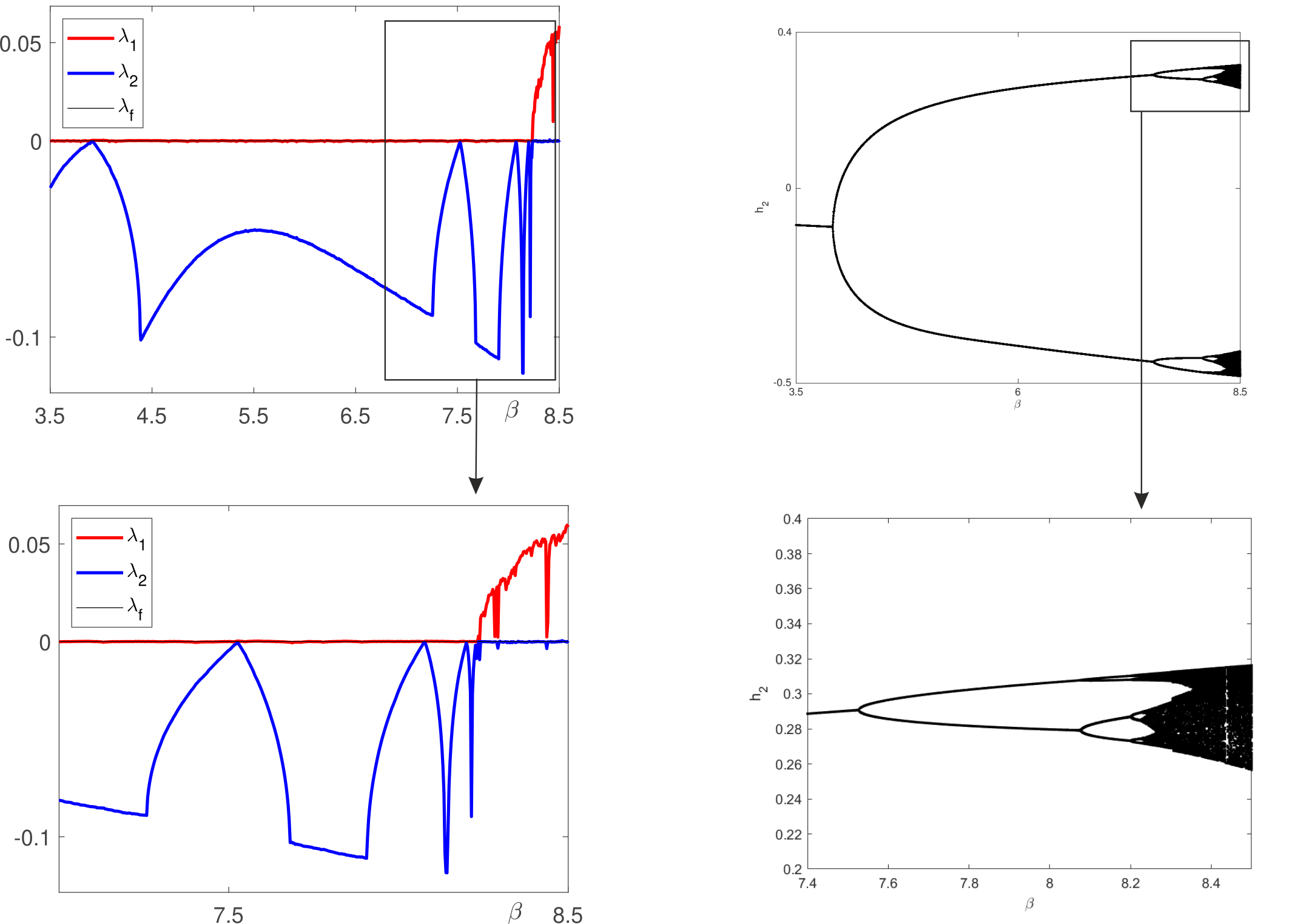}
     \caption{Lyapunov spectra and bifurcation trees for the Feigenbaum cascade of period doubling in \eqref{eq:Potts3_1} at $c=0.55$ and  $\beta\in[3.1,8.6]$.}
     \label{fig:fig8}
 \end{figure}

We demonstrate realization of this scenario in \eqref{eq:Potts3_1} at $c=0.75$ and $\beta\in[3.4,22]$. In Fig. \ref{fig:fig5} we show the Lyapunov spectra and the bifurcations trees for this region. From Fig. \ref{fig:fig5}B one can clearly see that, after one period-doubling bifurcation, a periodic attractor becomes quasiperiodic  one, undergoing the supercritical Neimark--Sacker bifurcation. We can observe a narrow but distinct region of the existence of quasiperiodic dynamics in Fig.  \ref{fig:fig5}B,C. Then this quasiperiodic orbit becomes resonant and a long-periodic orbit is born. Then this resonant periodic orbit becomes chaotic after undergoing a cascade of period doubling bifurcations and a chaotic attractor is born on the basis of former quasi-periodic orbit. We demonstrate Poincar\'e sections of the attractors along the line of the Afraimovich--Shilnikov bifurcation scenario in Fig. \ref{fig:fig6}.

We also show the dependence of the entropy production for the Potts model on the parameter $\beta$ and compare it with those of the Lyapunov spectrum for \eqref{eq:Potts3_1} (see, Fig. \ref{fig:fig7} ).
The average rate of entropy production can be calculated from the out-of-equilibrium definition of work for feedback-driven systems \cite{sagawa2012nonequilibrium,de2019oscillations}, we have for the feedback Potts model the formula
\begin{equation}
\sigma = \sum_a \dot{h}_a x_a
\end{equation}
From Fig. \ref{fig:fig7} one can see that the changes in the entropy production are in direct correlation with the bifurcations in the feedback model.

 \begin{figure*}[!ht]
    \centering
     \includegraphics[width=0.8\textwidth]{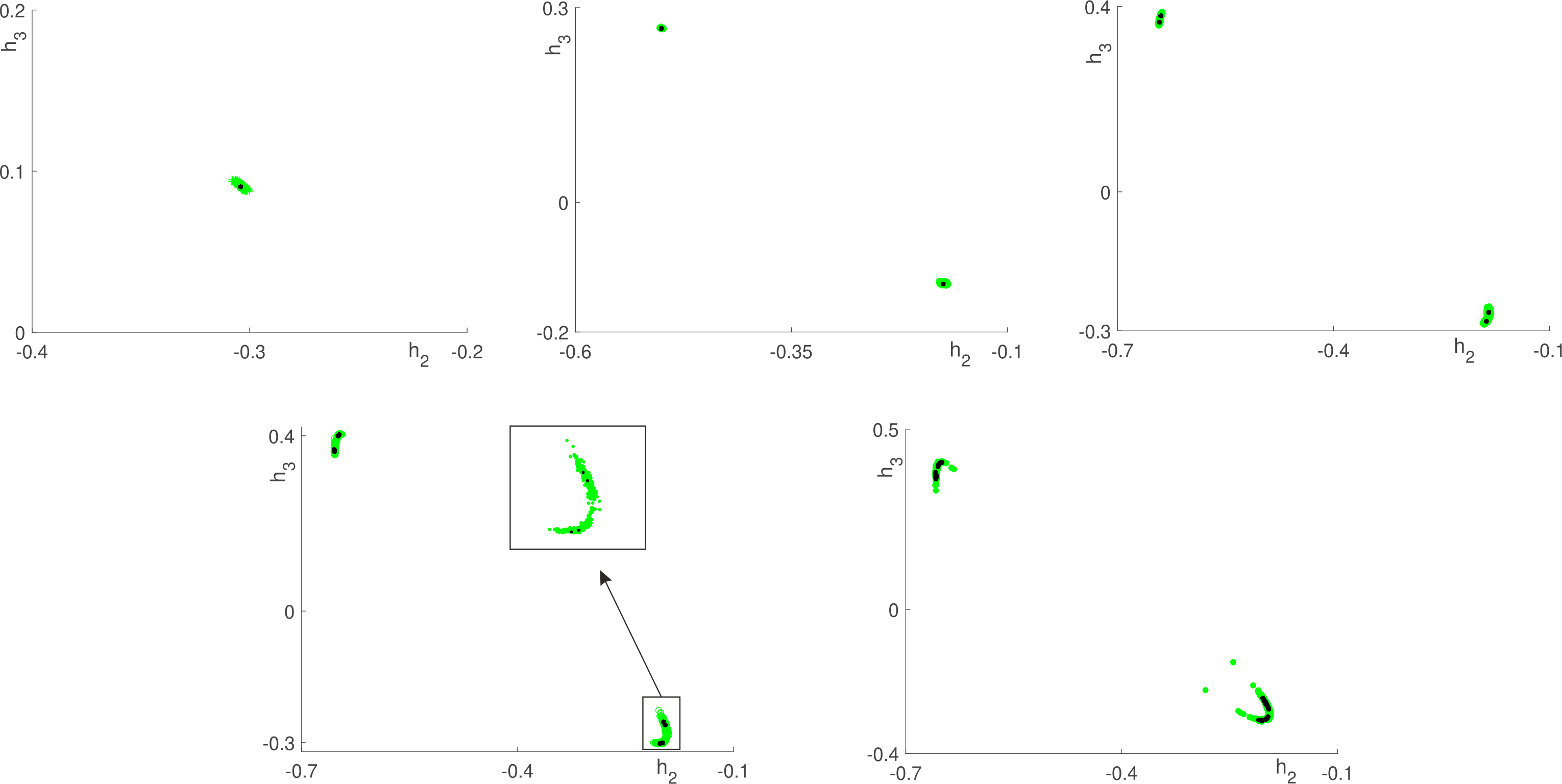}
     \caption{Poincare sections of the attactors that appear along the line of the Feigenbaum cascade of period doubling. In all plates $c=0.55$ and $\beta$ is $3.5$, $4.5$, $7.7$, $8.1$ and $8.3$, respectively. Numerical solutions of \eqref{eq:Potts3_1} are given in black, while green dots are obtained from fully microscopic stochastic simulations of a system of size $N=10^6$ spins via the Metropolis--Hastings method.}
     \label{fig:fig9}
 \end{figure*}

Chaotic attractors in \eqref{eq:Potts3_1} can also appear as a result of the Feigenbaum cascade of period doubling. This is typical for the lower part of the bifurcation chart in Fig.\ref{fig:fig4}. We suppose that $c=0.55$ and $\beta\in[3.1,8.6]$ and present the corresponding graphs of largest Lyapunov exponents and bifurcation trees in Fig.\ref{fig:fig8}. One can observe a typical cascade of period doubling which a periodic orbit undergoes. We demonstrate Poincar\'e sections of the attractors along the line of the Feigenbaum scenario in Fig. \ref{fig:fig9}. One can see that period doubling can be observed both in dynamical system \eqref{eq:Potts3_1} and in the microscopic stochastic simulations of a system of size $N=10^6$ spins.

\section{Conclusion}
In this work we have analyzed classical spin lattice models, namely the Ising model, the Blume--Capel model and the Potts model in presence of a negative feedback between the order parameter and the external field(s). These models are representative of the universality classes of equilibrium phase transitions  and in this work we have showed that their usual critical points and phase transition lines get transformed onto more complex bifurcations with the emergence of periodic, quasiperiodic and chaotic oscillatory patterns. At odds with the case of driven systems \cite{zhang2016critical,buendia2008dynamic,robb2014extended,mendes1991dynamics,keskin2005dynamic} these oscillations are  emerging self-oscillations and the system is autonomous, with no explicit dependence on time.

Our first general result is the derivation from linear response theory of simple lower dimensional systems of  differential equations  that quantitatively reproduce the many body stochastic simulations for fully connected models. These systems of equations can then be analyzed by classical tools of bifurcation theory.

In some cases the system inherits the main features of its equilibrium counterpart, including analytical tractability. This is the case of the Ising model, where we have shown that in finite dimensions, namely 2D, self-oscillations can emerge with a non-trivial exponent for the amplitude $\beta=1/8$ \footnote{As usual this shall not be confused with the inverse temperature.} in line with the celebrated Onsager solution for the static system.

We have demonstrated that  for the Blume--Capel model on a fully connected graph the usual tricritical point gets transformed into the Bautin bifurcation point and the second and first order phase transition lines are taken over by sub-critical and super-critical Andronov--Hopf bifurcation lines, respectively.  These results are independent of the feedback strength parameter $c$, as soon as this does not break the validity of the continuous approximation, that is valid in the thermodynamic limit and they are nicely summarized  qualitatively by the Landau theory of an homogeneous self-compatible field in presence of a feedback.

On the other hand we have shown that for the case of the Potts model with feedback, the dynamical picture is much more complex, due to the increased effective  dimensions in which the order and control parameters exist (from two to four in going from the Ising model to the Potts model with  $q=3$ colors).  The character of the bifurcation where self-oscillations emerge that substitutes the equilibrium phase transition (at $\beta_c=q$)  depends on the strength of the feedback $c$. For low enough $c$ it is discontinuous (sub-critical) and there exists a region where a stable fixed point coexists with self-oscillations. This corresponds well with the static equilibrium transition, that is known to be first-order for the fully connected Potts model at $q=3$.

However, if one increase $c$, the amplitude of the emerging self oscillations decreases up to a certain critical point where it becomes continuous akin to the Bautin bifurcation. This time it depends on the feedback strength and at odds with respect to the underlying free energy landscape. Furthermore, if one increases $\beta$ (i.e. decreases temperature), we have obtained that the bifurcation diagram of the Potts model with feedback endows complex scenarios with cascades of bifurcations leading to new limit cycles and quasiperiodic attractors and eventually to chaotic ones. The out-of-equilibrium thermodynamic features of these systems have been worked out numerically and we have demonstrated how singularities of the entropy production correspond to qualitative changes in the spectrum of the Lyapunov exponents and the underlying bifurcations.

Among the many future directions of this work we believe it would be interesting to analyze feedback lattice models with generalized out-of-equilibrium Landau functional formalism  \cite{guislain2022nonequilibrium} extending it beyond the Ising systems as well as to apply our framework to data analysis of collective oscillations in natural systems \cite{tort2021attractor}, especially synchronization of neuronal systems \cite{lombardi2023statistical}.
%paralleling the one of paradigmatic chaotic dynamical systems,

\section*{Acknowledgements}
DDM, AP and MFA thank Prof. Enzo Marinari for support and nice discussions. DS is grateful to Alexei Kazakov and Natalia Stankevich for useful discussions.

\bibliographystyle{elsarticle-num}
\bibliography{ref_osc}%

\end{document}